\documentclass[12pt,a4paper]{iopart}
\usepackage{iopams}  
\usepackage{bits} 
\usepackage{graphicx}
\usepackage{float}
\usepackage[breaklinks=true,colorlinks=true,linkcolor=blue,urlcolor=blue,citecolor=blue]{hyperref}

\newcommand{\isotope}[2]{$^{#2}{\rm #1}$}

\begin{document}

\title[Project~8]{Determining the neutrino mass with Cyclotron Radiation Emission Spectroscopy -- Project 8}


\author{
Ali~Ashtari~Esfahani$^1$,
David\,M.~Asner$^2$,
Sebastian~B\"{o}ser$^{3\dagger}$,
Raphael~Cervantes$^1$,
Christine~Claessens$^3$,
Luiz~de~Viveiros$^4$,
Peter\,J.~Doe$^1$,
Shepard~Doeleman$^5$,
Justin\,L.~Fernandes$^2$,
Martin~Fertl$^1$,
Erin\,C.~Finn$^2$,
Joseph\,A.~Formaggio$^6$,
Daniel~Furse$^6$,
Mathieu~Guigue$^2$,
Karsten\,M.~Heeger$^7$,
A.\,Mark~Jones$^2$,
Kareem~Kazkaz$^8$,
Jared\,A.~Kofron$^1$,
Callum~Lamb$^1$,
Benjamin\,H.~LaRoque$^4$,
Eric~Machado$^1$,
Elizabeth\,L.~McBride$^1$,
Michael\,L.~Miller$^1$,
Benjamin~Monreal$^4$,
Prajwal~Mohanmurthy$^6$,
James\,A.~Nikkel$^7$,
Noah\,S.~Oblath$^{2,6\ddagger}$,
Walter\,C.~Pettus$^1$,
R.\,G.\,Hamish~Robertson$^1$,
Leslie\,J.~Rosenberg$^1$,
Gray~Rybka$^1$,
Devyn~Rysewyk$^6$,
Luis~Salda\~{n}a$^7$,
Penny\,L.~Slocum$^7$,
Matthew\,G.~Sternberg$^1$,
Jonathan\,R.~Tedeschi$^2$,
Thomas~Th\"{u}mmler$^9$,
Brent\,A.~VanDevender$^2$,
Laura\,E.~Vertatschitsch$^5$,
Megan~Wachtendonk$^1$,
Jonathan~Weintroub$^5$,
Natasha\,L.~Woods$^1$,
Andr\'{e}~Young$^5$,
Evan\,M.~Zayas$^6$
}

\ead{$^{\dagger}$sboeser@uni-mainz.de, $^{\ddagger}$noah.oblath@pnnl.gov}

\address{$^1$ Center for Experimental Nuclear Physics and Astrophysics and Department of Physics, University of Washington, Seattle, WA, USA 98195}
\address{$^2$ Pacific Northwest National Laboratory, Richland, WA USA, 99354}
\address{$^3$ Institut f\"ur Physik, Johannes-Gutenberg Universit\"at Mainz, Germany, 55099}
\address{$^4$ Wright Laboratory and Department of Physics, University of California, Santa Barbara, CA, USA 93106}
\address{$^5$ Harvard-Smithsonian Center for Astrophysics, Cambridge, MA, USA 02138}
\address{$^6$ Laboratory for Nuclear Science, Massachusetts Institute of Technology, Cambridge, MA, USA 02139}
\address{$^7$ Department of Physics, Yale University, New Haven, CT, USA 06520}
\address{$^8$ Lawrence Livermore National Laboratory, Livermore, CA, USA 94550}
\address{$^9$ Institut f\"ur Kernphysik, Karlsruher Institut f\"ur Technologie, Karlsruhe, Germany 76021}

\begin{abstract}
The most sensitive direct method to establish the absolute neutrino mass is observation of the endpoint of the tritium beta-decay spectrum. Cyclotron Radiation Emission Spectroscopy (CRES) is a precision spectrographic technique that can probe much of the unexplored neutrino mass range with $\mathcal{O}({\rm eV})$ resolution. A lower bound of $m(\nu_e) \gtrsim 9(0.1)\un{meV}$ is set by observations of neutrino oscillations, while the KATRIN Experiment -- the current-generation tritium beta-decay experiment that is based on Magnetic Adiabatic Collimation with an Electrostatic (MAC-E) filter -- will achieve a sensitivity of $m(\nu_e) \lesssim 0.2\un{eV}$. The CRES technique aims to avoid the difficulties in scaling up a MAC-E filter-based experiment to achieve a lower mass sensitivity. In this paper we review the current status of the CRES technique and describe Project~8, a phased absolute neutrino mass experiment that has the potential to reach sensitivities down to $m(\nu_e) \lesssim 40\un{meV}$ using an atomic tritium source.
\end{abstract}


\section{Motivation}
The direct measurement of the absolute neutrino mass scale is one of the most pressing challenges in modern physics~\cite{Tribble:2007fk,Geesaman:2015aa}.  Several probes are used to measure the absolute neutrino mass, including cosmological measurements~\cite{Planck:2016,Lesgourgues:2014fk}, orbital electron capture processes~\cite{Alpert:2015,Gastaldo:2014}, and beta decay processes.  The tritium endpoint method, which relies on the tritium beta-decay, is both extremely sensitive to the neutrino mass and model independent~\cite{Otten:2008zz}.  This method is sensitive to neutrino mass via distortions imposed by energy conservation on the energy spectrum of electrons from tritium beta decay.  The nuclear matrix element and Coulomb correction for tritium beta decay are independent of neutrino mass, so the kinematic phase space factor alone determines the neutrino mass dependence of the spectral shape.  In addition, the neutrino mass determined from tritium decay is independent of whether neutrinos are Majorana or Dirac particles.  The current generation tritium endpoint experiment is the Karlsruhe Tritium Neutrino (KATRIN) experiment~\cite{Angrik:xw}. KATRIN will exhaust the quasi-degenerate range of neutrino masses that are large compared to either of the mass differences measured in oscillation experiments, or measure the neutrino mass if it lies in that range. Either outcome will determine whether neutrinos played a significant role in the formation of large scale structures in the universe~\cite{Hannestad:2005ey}. 

Neutrino oscillations do not provide any information about the absolute values of the neutrino mass eigenstates, but they provide a lower limit on the mass values of the second and third heaviest state. While the ultimate interest is to exhaust the entire range of masses allowed by oscillation bounds, the goal of the next generation of tritium endpoint experiments should be to reach the full range of neutrino masses allowed under an inverted hierarchical ordering of neutrino mass eigenvalues.    Figure~\ref{fig:hierarchies} shows the allowed neutrino masses in each ordering together with the projected 90\% confidence level from future  experiments~\cite{Angrik:xw,Lesgourgues:2014fk}. In this paper we present Project~8, a next-generation tritium endpoint experiment with a sensitivity as small as 40\un{meV} (90\% c.l.)~\cite{Doe:2013fk}, probing the full range of neutrino masses allowed by an inverted ordering.

\onefig[tbp]
  {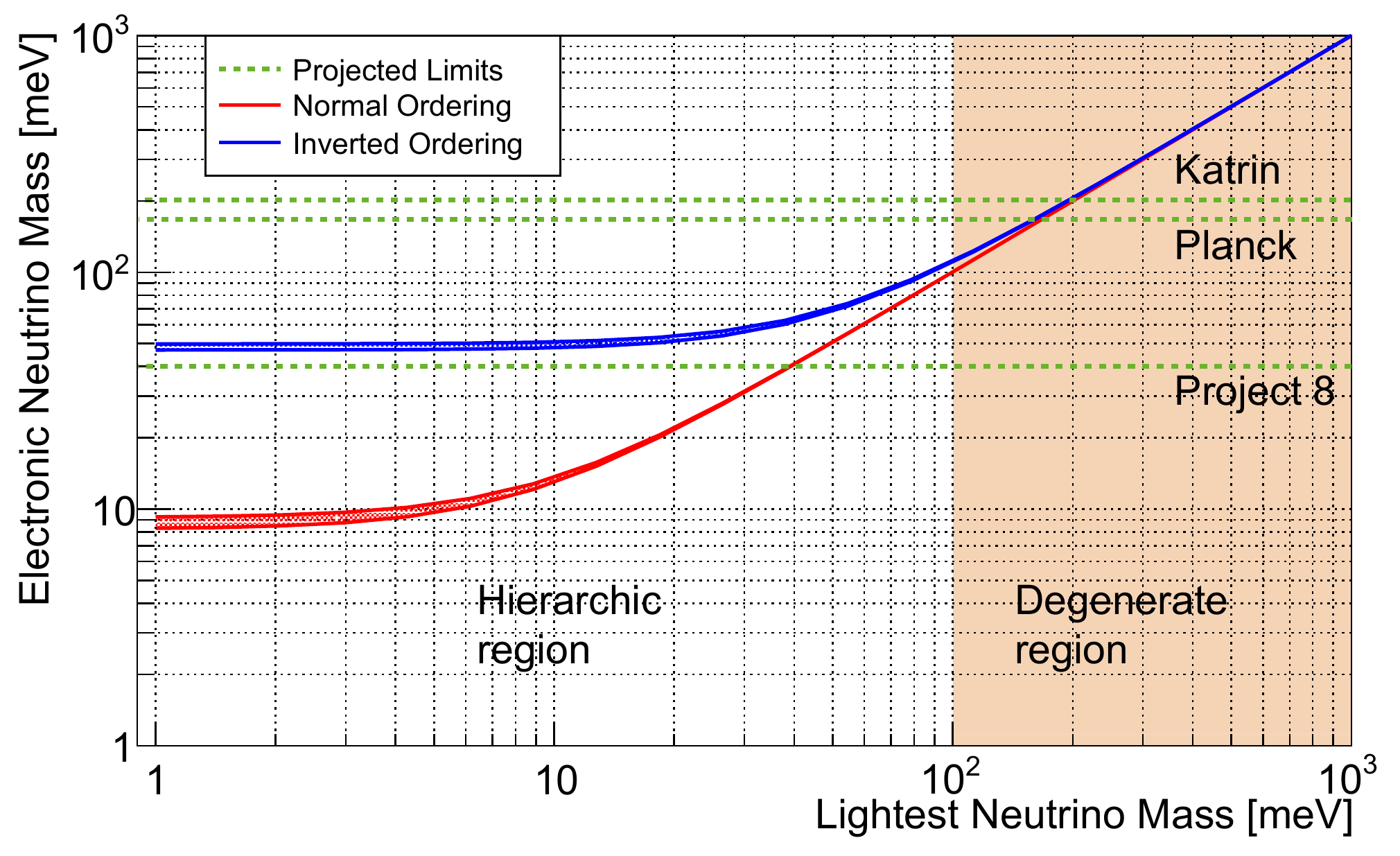}
  {\label{fig:hierarchies} Allowed range for the electron neutrino mass constrained by measured mixing parameters versus the value of the lightest mass eigenvalue. The green dashed lines show the projected 90\% c.l. electron neutrino mass limits from selected experiments.  Planck limits are from~\cite{Planck:2016}}

For this purpose, the Project~8 Collaboration has developed the technique of Cyclotron Radiation Emission Spectroscopy (CRES) to measure the electron spectrum near the tritium endpoint.  In the ideal CRES setup, a gaseous beta or conversion-electron source decays in a uniform magnetic field.  The electrons execute cyclotron motion, radiating power because of their centripetal acceleration.  A tritium endpoint electron with 18.6\un{keV} kinetic energy in a $\sim\!1$\un{T} magnetic field radiates about 1\un{fW} of cyclotron power at approximately 26\un{GHz}; the exact cyclotron frequency is directly related to the electron energy, and this relation depends only on the magnetic field strength and the electron mass and charge.  If the field is uniform and the cyclotron radiation can be observed for a few microseconds to determine its frequency, the extreme precision possible in the frequency domain translates to excellent energy resolution.  The axial motion of the electron can be assessed by observing the sideband frequencies that are a result of the Doppler shift of the emitted radiation.

\newcommand{\tableWidthRnD}{1.75in}
\newcommand{\tableWidthScience}{1.8in}

\setlength{\tabcolsep}{5pt}
\begin{table}
\caption{\label{tab:P8Phases} Phases of the Project~8 experiment.}
\begin{center}
\begin{tabular}{c|c|c|c|c}
\hline \hline \noalign{\smallskip}
\hspace{-.5ex}{\bf Phase} & {\bf Timeline} & {\bf Source} & {\bf R\&D Milestones} & {\bf Science Goals} \\
\noalign{\smallskip} \hline \noalign{\smallskip}
I & 2010--2016 & \isotope{Kr}{83{\rm m}} & \parbox[m]{\tableWidthRnD}{Single electron detection \\ Proof of concept} & \parbox[m]{\tableWidthScience}{Conversion electron \\ spectrum of \isotope{Kr}{83{\rm m}}} \\[2.9ex]
II & 2015--2017 & T$_2$ & \parbox[m]{\tableWidthRnD}{Kurie plot \\ Systematics studies} & \parbox[m]{\tableWidthScience}{Final-state spectrum test\\ \isotope{H}{3}$-$\isotope{He}{3} mass difference \\ $m_{\nu} \lesssim 10$--100\un{eV}} \\[4.5ex]
III & 2016--2020 & T$_2$ & \parbox[m]{\tableWidthRnD}{High-rate sensitivity \\ $B$ field mapping} & \parbox[m]{\tableWidthScience}{$m_{\nu} \lesssim 2$\un{eV}} \\[2.5ex]
IV & 2017-- & T &  \parbox[m]{\tableWidthRnD}{Atomic tritium source} & \parbox[m]{\tableWidthScience}{$m_{\nu} \lesssim 40$\un{meV} \\  Measure $m_{\nu}$ or determine normal hierarchy} \\[3.5ex]
\noalign{\smallskip}
\hline \hline
\end{tabular}
\end{center}
\end{table}
Project~8 is proceeding with the four-phase approach in Table~\ref{tab:P8Phases} to develop the next-generation tritium endpoint experiment based on CRES.
Each phase has distinct scientific goals and critical engineering milestones necessary to reach the sensitivity of $m(\nu_e) \lesssim 40\un{meV}$. The phases are broadly defined and will be conducted in parallel whenever possible.  Phase I has demonstrated the CRES technique~\cite{Asner:2015fk} originally proposed by Monreal and Formaggio~\cite{Monreal:2009za}.  A harmonic magnetic bottle ($B \sim z^2$) inside a rectangular microwave waveguide confined conversion electrons from \isotope{Kr}{83{\rm m}}.  The demonstration has been completed by replacing the harmonic trap with a wider ``bathtub'' trap allowing electron energy resolution $\Delta E \approx 1$\un{eV}. The distinguishing feature of Phase II is the first CRES measurement of tritium decay.  Phase II will occur inside a larger gauge circular waveguide for increased source volume and to reclaim the half of the cyclotron power lost by a rectangular waveguide.  Phase II data will test modern calculations of the molecular final-state spectrum~\cite{Bodine:2015aa}, measure the molecular tritium endpoint with $\lesssim 10$\un{eV} precision, and set a neutrino mass limit $m_\nu \lesssim 10\un{-}100$\un{eV}.  Phase III will eliminate the waveguide, instrumenting a tritium source in free space with a phased array of antennas.  This is a critical step towards the large volume required to accommodate enough tritium for a high event rate and sufficient statistical sensitivity.  Phase III will give a limit $m_\nu \lesssim 2$\un{eV}, competitive with current limits from the Mainz and Troitsk experiments~\cite{Kraus:2005fv,Aseev:2011fk}.  All of the tritium sources through Phase III will employ molecular tritium gas, denoted T$_2$.  Phase IV will include development and operation with {\em atomic} tritium (T) sources.  Atomic tritium is required to avoid an irreducible systematic uncertainty~\cite{Bodine:2015aa} associated with the final states of the \isotope{He}{3}T$^+$ ion populated by beta decay of T$_2$.  Phase IV will have sensitivity to the entire inverted hierarchy and to the normal hierarchy down to 40\un{meV}.

Section~\ref{sec:tritium_endpoint} of this document reviews the tritium endpoint method
to motivate our new spectroscopy method, CRES, which is
introduced in Section~\ref{sec:CRES}.  Section~\ref{sec:phaseI} reviews the results of Phase I, which provided a proof-of-principle.  Looking towards a neutrino mass measurement, Section~\ref{sec:sensitivity} covers the calculation of neutrino-mass sensitivity, Section~\ref{sec:phaseII} discusses current progress on the construction of a tritium demonstrator for Phase II. Section~\ref{sec:phaseIII} outlines design studies for a Phase III apparatus, Section~\ref{sec:phaseIV} gives a conceptual design for Phase IV's atomic tritium source and an estimate of the sensitivity of a complete atomic tritium experiment.


\section{The Tritium Endpoint Method\label{sec:tritium_endpoint}}

The most auspicious place to look for the absolute scale of neutrino masses is in the kinematics of tritium beta decay~\cite{Otten:2008zz}.  Defining $E_0$ as the maximum energy available to the electron in the case where the $m_\nu = 0$ and atomic electrons are not present, we introduce $\epsilon \equiv E_0 - E$ and find a simple form of the electron energy spectrum near its endpoint:

\begin{equation}\label{eqn:tritiumBetaDecay}
\frac{dN}{d\epsilon} = 3rt \epsilon \sqrt{\epsilon^2 - m_{\beta}^2}.
\end{equation} 
Here, $r$ is the rate in the last 1\un{eV} of the spectrum with $m_\nu =0$ and $t$ is the observation time.  The observable $m_\beta^2$ is defined in terms of the mass eigenvalues $m_i$ and mixing matrix elements $U_{ei}$:
\begin{equation}\label{eqn:effectiveMass}
 \ m_{\beta}^2 = \sum_{i=1}^3 \left| U_{ei} \right|^2 m_i^2.
\end{equation}
This is the effective mass of the electron flavor antineutrino (hereafter, ``the neutrino'' for brevity).  The tritium endpoint method measures this mass via a very precise tritium beta decay electron spectrum fit to the form of Equation~(\ref{eqn:tritiumBetaDecay}) with $m_\beta^2$ as a free parameter.

The simple form of Equation~(\ref{eqn:tritiumBetaDecay}) belies the extreme difficulty of a tritium endpoint experiment.  Besides the obvious requirement for very good energy resolution ($\Delta E \sim 1$\,eV at $E_0 = 18.6$\un{keV}), the statistical sensitivity has unfortunate scaling relationships and the natural form of tritium gas, molecular T$_2$, has an irreducible systematic associated with final states \cite{Bodine:2015aa}.  The first unfavorable scaling relation follows from the extreme rarity of events near the endpoint; only $2 \times 10^{-13}$ of all events occur in the last 1\un{eV} of the spectrum. Therefore, a large amount of tritium is required to record sufficiently many events near the endpoint, while the vast majority of the spectrum is useless. 
The form of Eq.~\ref{eqn:tritiumBetaDecay} means that for an order of magnitude improvement in sensitivity to $m$, we need an improvement of 4 to 5 orders of magnitude in statistical sensitivity alone--accompanied by commensurate improvements in systematics.  
The large size of current generation tritium endpoint experiments follows from the needs to accommodate sufficient tritium source intensity for statistical sensitivity at the spectrum endpoint and to manage the uninteresting low-energy events. The final state systematic reflects uncertainty in the width of a narrow band of rotational and vibrational states of the \isotope{He}{3}T$^+$ daughter, populated in the decay of molecular tritium. The spectrum of final states is an irreducible systematic uncertainty in any experiment with molecular tritium.  

Electron spectroscopy in current state-of-the-art tritium endpoint experiments is performed by magnetic adiabatic collimation with an electrostatic filter (MAC-E)~\cite{Lobashev:1985mu}. Tritium gas decays in a strong solenoidal magnetic field $B_s$. The decay electrons are transported along field lines in a spiraling motion to a region of low field $B_a$, adiabatically converting momentum components perpendicular to the field to parallel motion. The relative residual perpendicular component of the electron momentum $p_\perp$ is driven by the ratio of magnetic field strength ${p_\perp}/{p_\|} \sim {B_a}/{B_s}$ in the high- and low-field regions. The collimated electrons are selected by an electrostatic potential in the low-magnetic-field region. Electrons with enough energy to clear the potential barrier are reaccelerated on the other side and refocused into a second region of high magnetic field where they impact a focal plane detector. Lower energy electrons are reflected and returned to the source region. A MAC-E filter is therefore a high-pass energy filter measuring the integral electron energy spectrum above the threshold set by the potential barrier. The differential spectrum must be constructed by scanning the potential.  

The Mainz and Troitsk apparatuses used to set the current limit on neutrino mass were both MAC-E spectrometers. KATRIN also uses a MAC-E spectrometer, and will improve in statistical sensitivity primarily due to a much more intense tritium source. KATRIN's source will have the maximum tolerable column density consistent with the requirement to transport electrons to the MAC-E spectrometer without significant probability of scattering in the source. A source supplying any higher density of tritium along the direction of the magnetic field is not useful.  The statistical sensitivity can only be improved by enlarging the source radially, with proportional radial expansion of the MAC-E spectrometer. The neutrino mass sensitivity of a MAC-E spectrometer therefore scales with the area of the source. KATRIN's spectrometer is already 10\un{m} in diameter and maintained at $10^{-11}$\un{mbar} pressure.  An order of magnitude increase in sensitivity would require a spectrometer at least 300\un{m} in diameter at the same pressure.  The extreme scale of such an apparatus means MAC-E is likely not a viable path forward from KATRIN.
A new technique will be required to go beyond KATRIN's $0.2\un{eV}$ limit.

\section{Cyclotron Radiation Emission Spectroscopy\label{sec:CRES}}
In Cyclotron Radiation Emission Spectroscopy (CRES) the energy of an electron determines the frequency of the electromagnetic radiation the electron emits when gyrating in a magnetic field $B$. The orbital revolution frequency $f$ of the electron, called the cyclotron frequency, depends on the kinetic energy $E_{\rm kin}$ of the electron.

\begin{equation}\label{eqn:f_cyclotron}
f = \frac{f_0}{\gamma} = \frac{1}{2\pi} \frac{eB}{m_e + E_{\rm kin}/c^2},
\end{equation}
where $m_e$ is the electron mass, $c$ is the speed of light in vacuum and $\gamma = 1 + E_{\rm kin}/m_ec^2$ is the Lorentz factor. As long as the axial motion of the electron remains non-relativistic, the coherent electromagnetic radiation is strongly peaked at $f$, with a non-relativistic low-energy limit of $f_0 = 2.799 249 110 \times 10^{10}\un{Hz}$ in a $1\un{T}$ magnetic field.
Because of the dependence of $f$ on kinetic energy, a frequency measurement of this radiation is related to the energy of the electron and thus enables a new form of nondestructive spectroscopy.
Along with frequency-based techniques inherited from other precision experiments, this gives the advantage of highly accurate measurements with relatively straightforward engineering. For a given electron, the energy resolution $\Delta E/E \sim \Delta f/f$ is to first order given by the accuracy of the frequency measurement, which in turn is linked to the time $\tau \sim 1/\Delta f$ for which the electron is observed. At a kinetic energy of $E_{\rm kin} = 18.6\un{keV}$--as in the $\beta$-decay of tritium--achieving an energy resolution of $\sim 1\un{eV}$ therefore requires trapping and observing the electron for several microseconds. The gas pressure is optimized such that the typical track length is long enough to measure the initial frequency with high precision.

The minuscule total power $P$ radiated by each electron provides a challenge to measurement;
in free space it is given by the Larmor formula,
\begin{equation}
 P = \frac{2 \pi e^2 f_0^2}{3 \epsilon_0 c} \frac{\beta^2 \sin^2\theta}{1-\beta^2} \sim 10^{-15}\un{W},
\end{equation}
where $\epsilon_0$ is the permittivity of free space, $\beta = v_e/c$ is the  electron velocity and $\theta$ is the pitch
angle of the electron. This is the angle between the
momentum vector of the electron and the direction of
the magnetic field. For an electron with an energy near the
$18.6\un{keV}$ endpoint of T$_{2}$ approximately $1.2\un{fW}$ is radiated in a 1\un{T} magnetic field at a pitch angle of $90^\circ$. 

While this calls for low-noise sensing techniques, such a frequency-based technique has the capability of overcoming many of the limitations imposed
by traditional spectroscopic techniques used in direct
neutrino mass experiments. The most sensitive
methods in use today require extracting the
beta decay electron for measurement, limiting
the size and density of the tritium source.
Because tritium gas is transparent to cyclotron radiation, this
restriction does not apply to the detection of cyclotron radiation. 
CRES is simultaneously sensitive to the entire energy range of interest, and naturally provides event-by-event energy reconstructions, eliminating two shortcomings of the traditional stepped integration.
Furthermore, these reconstructions employ well-established techniques for
measuring frequencies and magnetic fields.

\section{Phase I: Proof-of-Principle\label{sec:phaseI}}
The Project~8 collaboration recently demonstrated the CRES technique~\cite{Asner:2015fk} and proposes to use it for a tritium endpoint experiment. Figure~\ref{fig:apparatus} shows the CRES prototype instrument and a diagram of its receiver electronics chain.
\subsection{Experimental setup}
The setup consists of a small rectangular waveguide cell in which low-pressure gaseous \isotope{Kr}{83\rm m} is confined by two 25$\un{\mu m}$-thick Kapton windows. The decay electrons travel in an axial magnetic field while emitting cyclotron radiation. The waveguide captures and transmits the microwave radiation to the input of a low-noise radiofrequency receiver and digitizer. \isotope{Kr}{83\rm m} is appropriate as a test gas since it provides several narrow electron lines with less then 3\un{eV} width--one at $17.83\un{keV}$ very close to the tritium $\beta$-spectrum endpoint and four more at $30.23\un{keV}$, $30.42\un{keV}$, $30.48\un{keV}$ and $31.94\un{keV}$ \cite{Picard:1992ys}--which enables verification of the measurement technique's linearity over a large energy range. A steady supply of \isotope{Kr}{83\rm m} with a half-life of $1.8\un{h}$ is generated by the decay of a \isotope{Rb}{83} source adsorbed on zeolite beads \cite{Venos:2005vn}, from which the krypton diffuses freely through the experimental system. Getter pumps reduce the pressure from non-noble gases to $<10\un{\mu Pa}$.

\onefig[h]{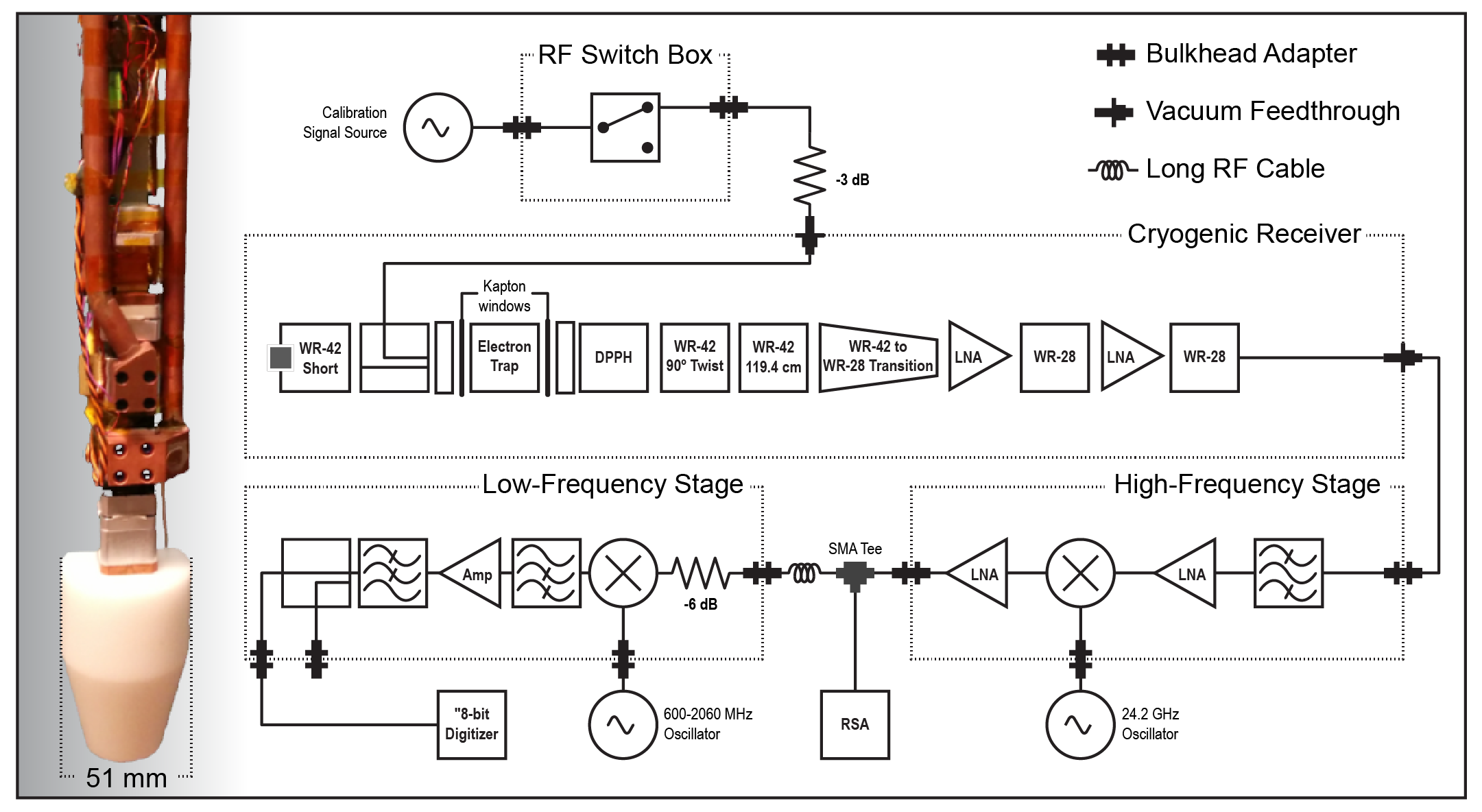}{\label{fig:apparatus} The prototype CRES instrument, and its receiver electronics, from~\cite{Asner:2015fk}.  The insert, which has a maximum diameter of 51\un{mm}, is placed in the 52\un{mm} bore of a solenoidal NMR magnet. The receiver chain consists of cascaded cryogenic amplifiers, a high-frequency stage, and a low-frequency stage. The frequency band of interest, with a width of 2\un{GHz} centered at 26\un{GHz}, is then mixed down to be centered at 1.8\un{GHz} with the same 2-GHz bandwidth and then either recorded or run through a second amplification and mixing stage.}

Phase I uses a warm-bore superconducting solenoid to supply an axial magnetic field; the 52\un{mm} bore diameter is the primary limit on the gas cell volume. 
The cell is placed in the center of the bore in a $\sim 1\un{T}$ magnetic field, which confines the electrons radially. On top of this uniform field, a weak magnetic
trap is added to confine the electrons axially and allow
sufficient time to detect and measure the cyclotron emission. This trapping
field is generated by a combination of three copper coils around the gas
cell that each provide near-harmonic field perturbations with a maximum depth
of $-8.2\un{mT}$ for a coil current of 2\un{A}. The resulting field gradient of up to
100\un{Tm^{-2}} along the magnetic field axis confines electrons with pitch angles between $85^{\circ}$ to $90^{\circ}$ relative to the magnetic field. In the {\it harmonic trap} configuration, only a
single coil is energized to decrease the magnetic potential in the center of the
cell volume, resulting in an effective source volume (the product of real
physical volume and magnetic trapping efficiency) of a few $\un{mm}^3$. In the {\it
bathtub trap} configuration, two coils at each end of the cell volume
generate a potential barrier for the electrons. When the same field gradient is used at the ends of the bathtub trap as were used in the above harmonic trap, electrons with the same range of pitch angles are trapped.

As the fundamental cyclotron
signals for the $30.4\un{keV}$ and $17.8\un{keV}$ electrons are expected to lie in the
microwave K band for a field strength of 1\un{T}, a standard WR42 waveguide is
used for the cell as well as to transport the signal towards the receiver. In
this co-axial configuration of the waveguide and the B-field, the cyclotron
emission couples strongly to the fundamental TE10 mode of the waveguide, and
a large fraction of the emitted power is detected. The cyclotron radius for a $17.8\un{keV}$ electron is $0.46\un{mm}$ in a 1~T magnetic field, which is small compared to the waveguide inner dimensions ($7.112\un{mm} \times 3.556\un{mm}$).

Two cascaded 22-40\un{GHz} low-noise preamplifiers form the first stage of the receiver chain with a gain of 54\un{dB}. At a physical temperature of 50\un{K}, this achieves an effective noise temperature of 145\un{K}. All later stages provide negligible contributions. For ease of digitization, the frequency band of interest around 26\un{GHz} is first mixed down with a local fixed-frequency 24.2\un{GHz} oscillator. A second mixer with a variable local oscillator frequency combines with a low-pass filter to select a frequency subband of 125\un{MHz} for narrow-band signal analysis. Signals are digitized at 250\un{MSPS} with a free-running 8-bit digitizer and recorded to disk. The low noise of the system ensures excellent signal fidelity with signal-to-noise ratio of 12\un{dB} for an 18\un{keV} electron and a receiver detection bandwidth of 30\un{kHz}.

\subsection{Results}

\onefig[tbh]{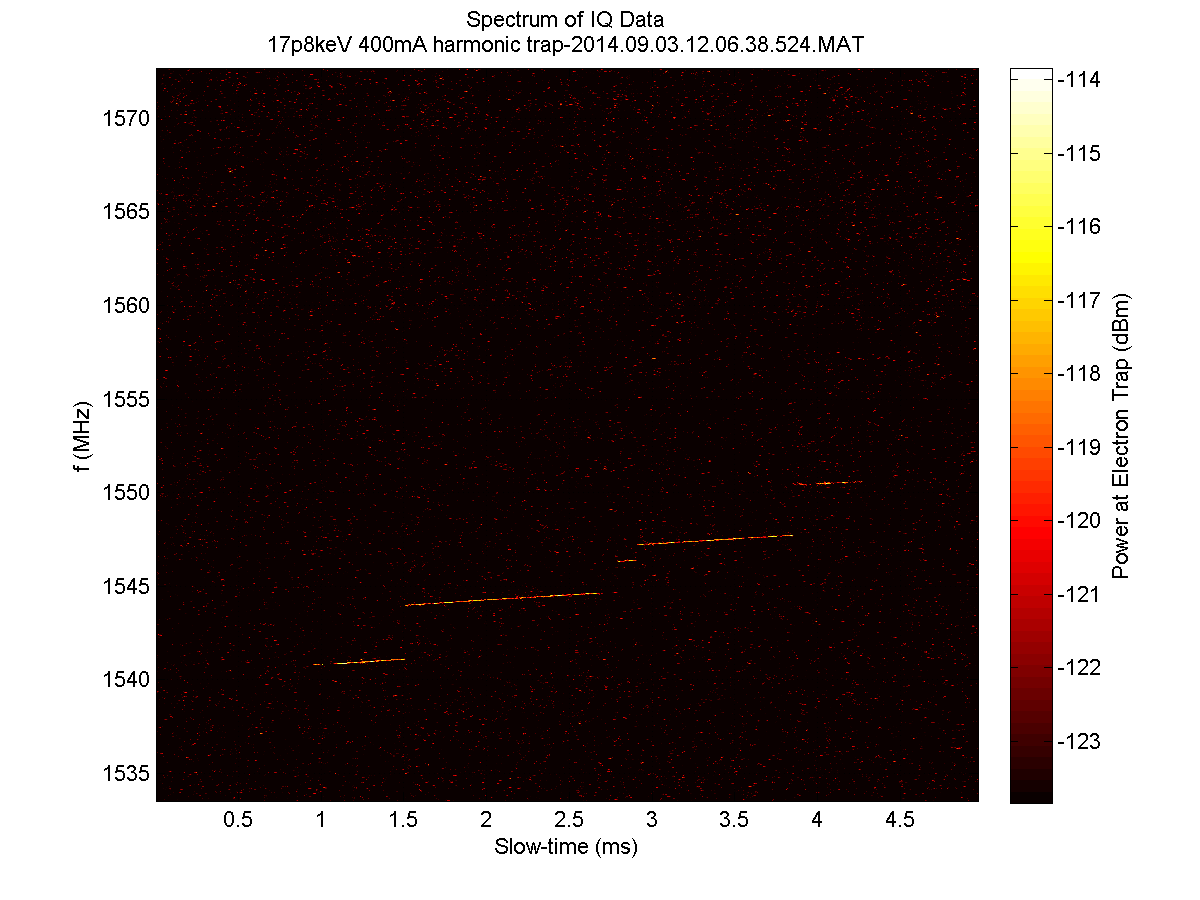}{\label{fig:waterfall} An individual \isotope{Kr}{83{\rm m}} conversion electron with 17.8\un{keV} observed by CRES. The frequency axis is the output after 24.2\un{GHz} down-conversion. The color represents power in dBm (-120\un{dBm} = 1\un{fW}).}

Figure~\ref{fig:waterfall} shows the spectrogram of a typical CRES event.
The event begins abruptly, chirps towards higher frequency (lower energy) as the
electron radiates cyclotron power, and makes a series of frequency (energy) hops
as it undergoes discrete scatterings on residual hydrogen in the high-vacuum environment.
To extract the initial energy of the electron, track segments are identified
using pattern recognition techniques on the spectrogram bins with a
signal-to-noise ratio of ${\rm SNR} > 6\un{dB}$. We associate all segments produced by one electron using the condition that a segment extending the current event must begin at the same time the previous segment ended. 
The initial energy of the electron is then derived from the
frequency at the starting point of the first track.

With this approach, an energy resolution of 16\un{eV} FWHM at the 30.4\un{keV} \isotope{Kr}{83{\rm m}} conversion electron line was achieved~\cite{Asner:2015fk} using a single coil to generate a coaxial parabolic
trapping field $B = B_0 {\boldsymbol{-}} \beta z^2$, with $B_0 = 0.94\un{T}$ and $\beta \sim
10^{-3}\un{T/mm^2}$. Reducing the trap depth, $\beta$, limits the pitch angle range under which
electrons are confined and therefore the rate of observable events; however it improves the energy resolution 
because it reduces the axial and radial variation in the magnetic field across the volume sampled by trapped electrons, 
and therefore it reduces the uncertainty on the average magnetic field experienced by any given electron.

Using the ``bathtub'' configuration $B = B_0 \boldsymbol{+} \beta_1 (z-z_1)^2
\boldsymbol{+} \beta_2
(z-z_2)^2 $, with two coils located at $z_1$ and $z_2$, results in a significant
improvement in energy resolution, as the average magnetic field variation is significantly smaller
over the intermediate range $z_1 < z < z_2$ probed by the trapped electrons. Figure~\ref{fig:spectrum} shows the energy spectra of
\isotope{Kr}{83{\rm m}} conversion electrons near the 30.4 and 32\un{keV}
lines.  To ensure we measure the genuine initial energy, we cut on the relative time of the event's beginning and the trigger time of the data acquisition system. 
With a selection efficiency of $>70\%$ and resulting FWHM of 3.3\un{eV}  for the two lines at 30.4\un{keV} and 3.6\un{eV} for 32\un{keV} lines, this gives a major resolution improvement over the initial configuration~\cite{Asner:2015fk}. We are studying the dependence of observed power of the cyclotron emission on electron energy, and should be able to improve the energy resolution such that it approaches the natural line width.

The data from Phase I of Project~8 will also permit the first quantification of CRES background rates.
There is only one known quantifiable background: the ejection of delta electrons
by cosmic rays passing through the source gas. This background rate has been calculated for
the KATRIN experiment~\cite{Leber:2005aa} and depends on the source density and
volume. Even for an effective volume of 1000\un{m^3} and a source density of
$2\times10^{13}$ molecules per \un{cm^{3}}, the background event rate is smaller
than $10^{-7}\un{(eV\cdot s)^{-1}}$.

\twofigone[tbh]
  {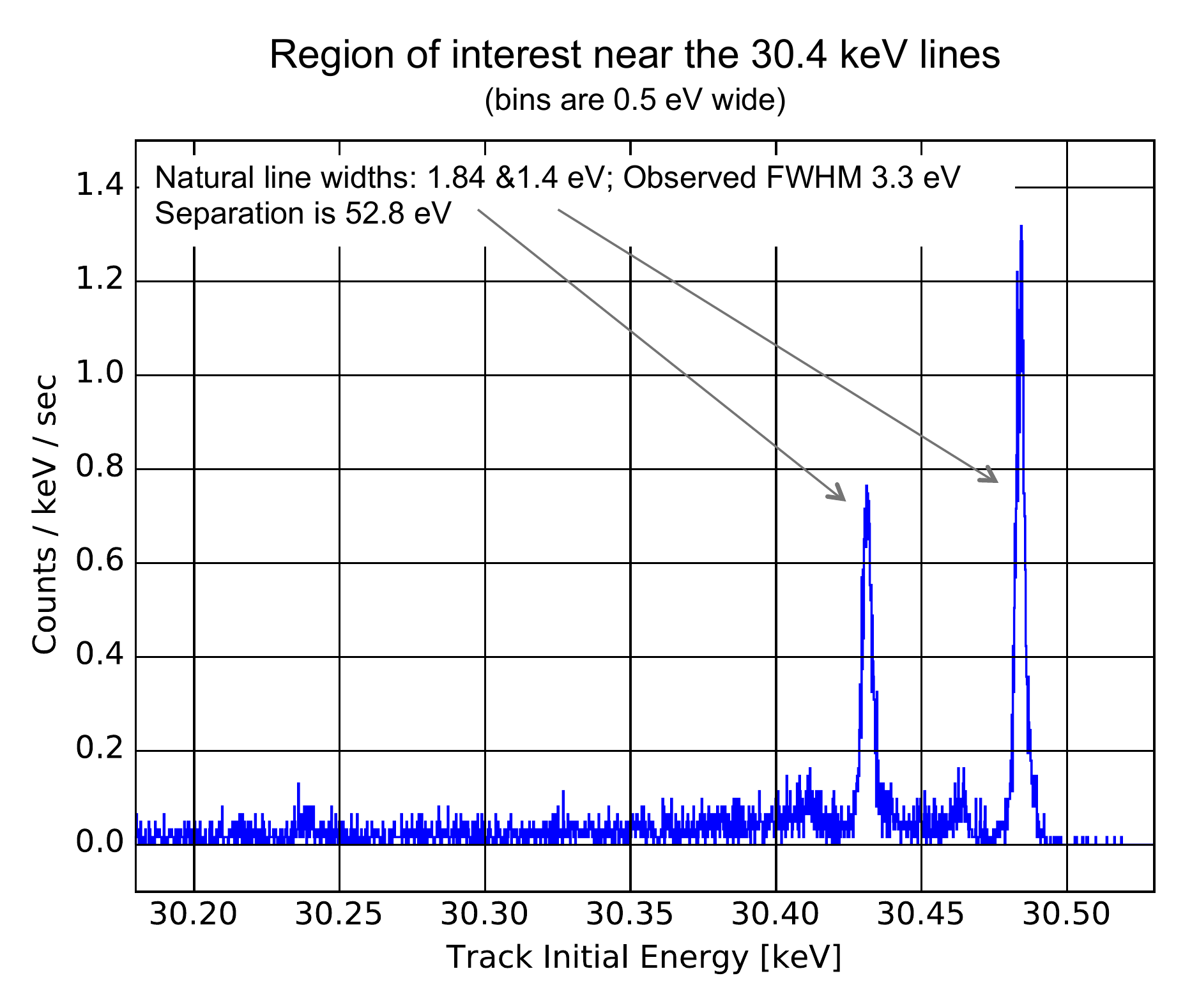}
  {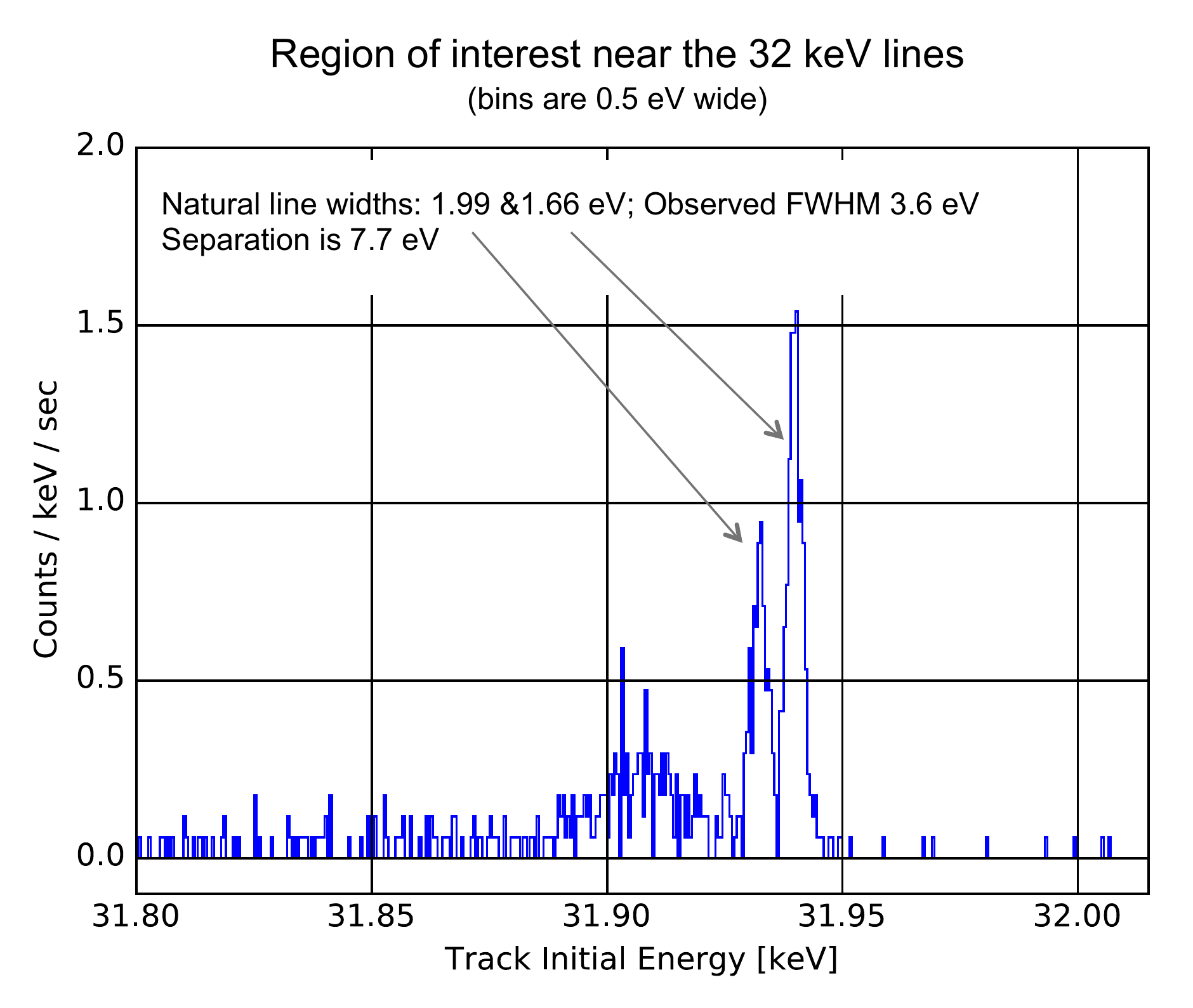}
  {Energy spectrum of the \isotope{Kr}{83{\rm m}} lines at 30.4\un{keV} (left)
    and 32\un{keV} (right) in the ``bathtub''
  configuration. The data shown here were recorded with a Tektronix 5106b Realtime Spectrum Analyzer, and analyzed in a manner similar to the procedure used in~\cite{Asner:2015fk}.\label{fig:spectrum}} 

\section{Towards a neutrino mass experiment\label{sec:tonumass}}
\subsection{Neutrino mass sensitivity\label{sec:sensitivity}}

With a successful proof-of-concept that demonstrates not only the ability to
detect the cyclotron radiation of single electrons but also the spectroscopic
resolution required for a neutrino mass experiment, Phase I of the Project~8 experiment
concluded in spring 2016.  With Phase II of the experiment, the collaboration is progressing 
towards a neutrino mass measurement.

The statistical sensitivity for an observation interval $\epsilon$ below $E_0$ is estimated~\cite{Doe:2013fk} to be
\begin{equation}\label{eqn:statistical_sensitivity}
\sigma_{m_{\nu}^2}^{\rm stat} = \frac{2}{3rt} \sqrt{rt \epsilon + \frac{bt}{\epsilon}},
\end{equation}
where $r$ is the rate in the last 1\,eV of the $m_\nu = 0$ spectrum, $t$ is the observation time, and $b$ is the background rate.  Contributions from molecular daughter ion final states, magnetic field uncertainty and thermal Doppler and collisional broadening ({\it i.e.}, finite mean free lifetime) are also considered.  For concreteness, an assumption is made that contributions from final states and collisional broadening are or will be known to 1\%.  The magnetic field effect is assigned an RMS magnitude of $10^{-7}$.  The assumed background rate is $b = 10^{-6}\,({\rm eV \cdot s})^{-1}$.  Figure~\ref{fig:P8sensitivity} shows how the 90\% confidence sensitivity scales with the effective volume of the experiment, for four different source scenarios and one year of cumulative observation time. The effective volume, which includes instrumental efficiencies, could be as low as 10\% due to the requirement for magnetic trapping and the $\sin^2\theta$ dependence of cyclotron power in Equation~(\ref{eqn:f_cyclotron}).  The source scenarios are denoted as molecular (T$_2$) or atomic (T) tritium and the number density of source molecules (or atoms) is in units of $\un{cm}^{-3}$.  The statistical sensitivity and final state spectrum are discussed above, and the background $b$ is assumed to be constant per unit energy for Project~8.  The sensitivity improves with increasing effective volume ({\it i.e.}, increased statistics associated with total source strength) up to a plateau in all scenarios.  The highest plateau is for the densest molecular tritium source.  The limit in that case is collisional broadening--the mean free path for electrons is not long enough for a sufficient determination of frequency.  For molecular sources at lower densities, collisional broadening is replaced by the final state spectrum of the the daughter \isotope{He}{3}T$^+$ ion as the limiting factor.  This sensitivity is similar to that of KATRIN and is the best that can be done with a molecular source.  For an atomic source at 1\,K and sufficiently low density to allow long mean free paths, the sensitivity can be further improved, approaching 40\un{meV} in a large instrument.  The limit is caused by assumed magnetic field uncertainty $\Delta B / B \sim 10^{-7}$.

In addition to these analytical sensitivity calculations, an analysis of the factors expected to influence Project~8 sensitivity is being performed using the Stan Markov Chain Monte Carlo simulation package~\cite{Carpenter:2015aa, Stan:2015aa}. Stan allows for Monte Carlo-based statistical modeling and simultaneous optimization of a large number of parameters. We simulate beta decay spectra assuming either a normal or an inverted neutrino mass hierarchy, then analyze these simulated spectra given models of spectral shape, endpoint energy distribution, and Project~8 background rate and instrumental efficiencies. The spectra are generated and analyzed for a source with variable composition fractions of T$_2$, HT, DT, and T, so it is possible to account for source contamination and compare the projected systematics of Phases I-IV.

\onefig[H]{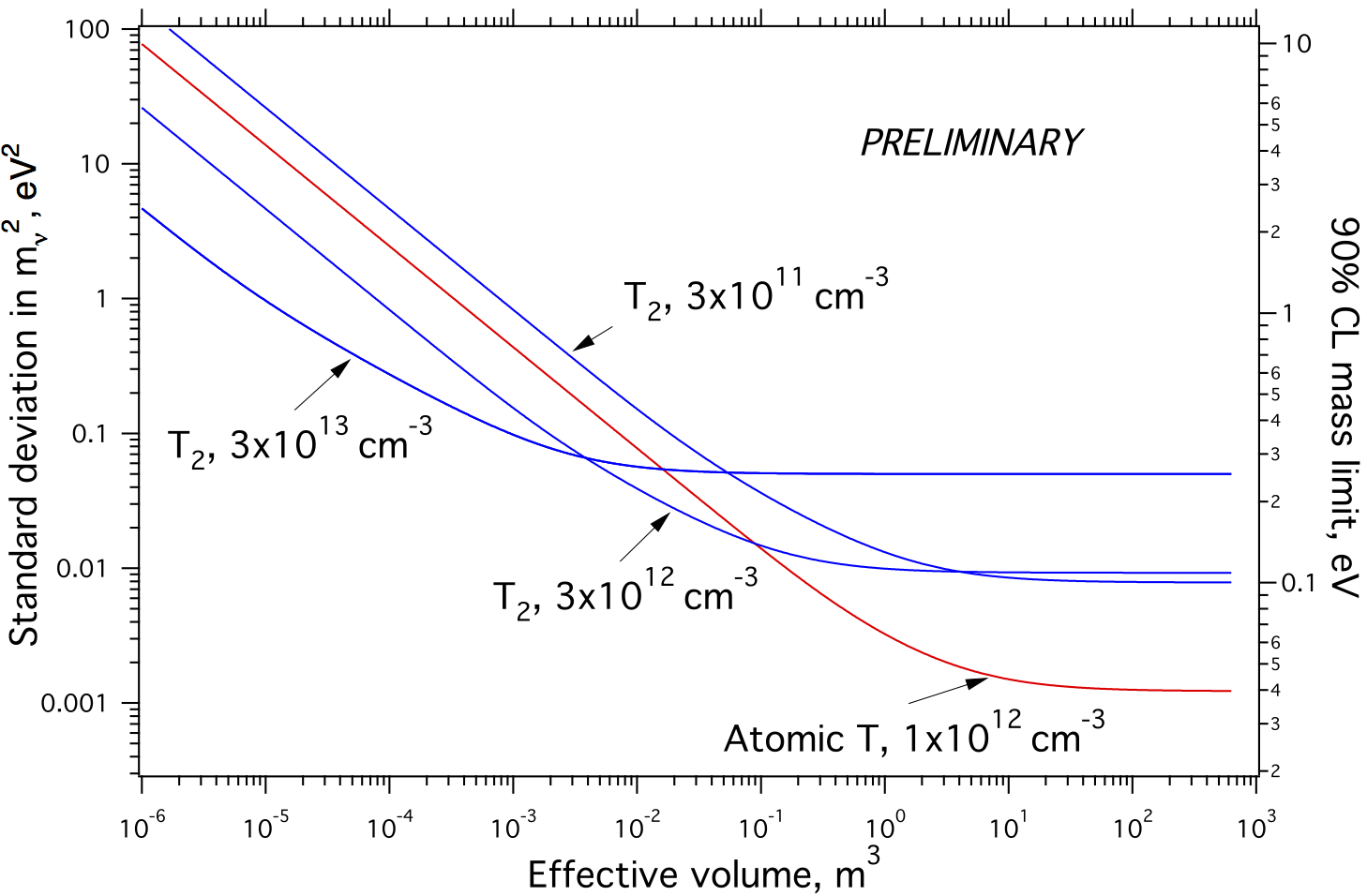}{\label{fig:P8sensitivity} Projected standard deviation (left axis) and the corresponding neutrino mass limit (right axis) of Project~8 under several scenarios.  Each curve is labeled with a number density, and either T or T$_2$ for atomic or molecular sources, respectively.  All curves assume one year of cumulative run time.}

\subsection{Phase II: Tritium demonstrator\label{sec:phaseII}}
This phase marks the first investigation of tritium by the CRES technique, on a
small scale similar to Phase I. A tritium compatible cell has been built and
will be charged with a low pressure of T$_2$ gas. Although the effective volume
remains small, we will nevertheless be able to obtain significant
scientific results from this apparatus. The variance of the final-state spectrum
can be measured for comparison with theory at the few percent level, and a
measurement made of the atomic mass difference between \isotope{He}{3} and
tritium at a competitive level.

\onefig[tbh]
  {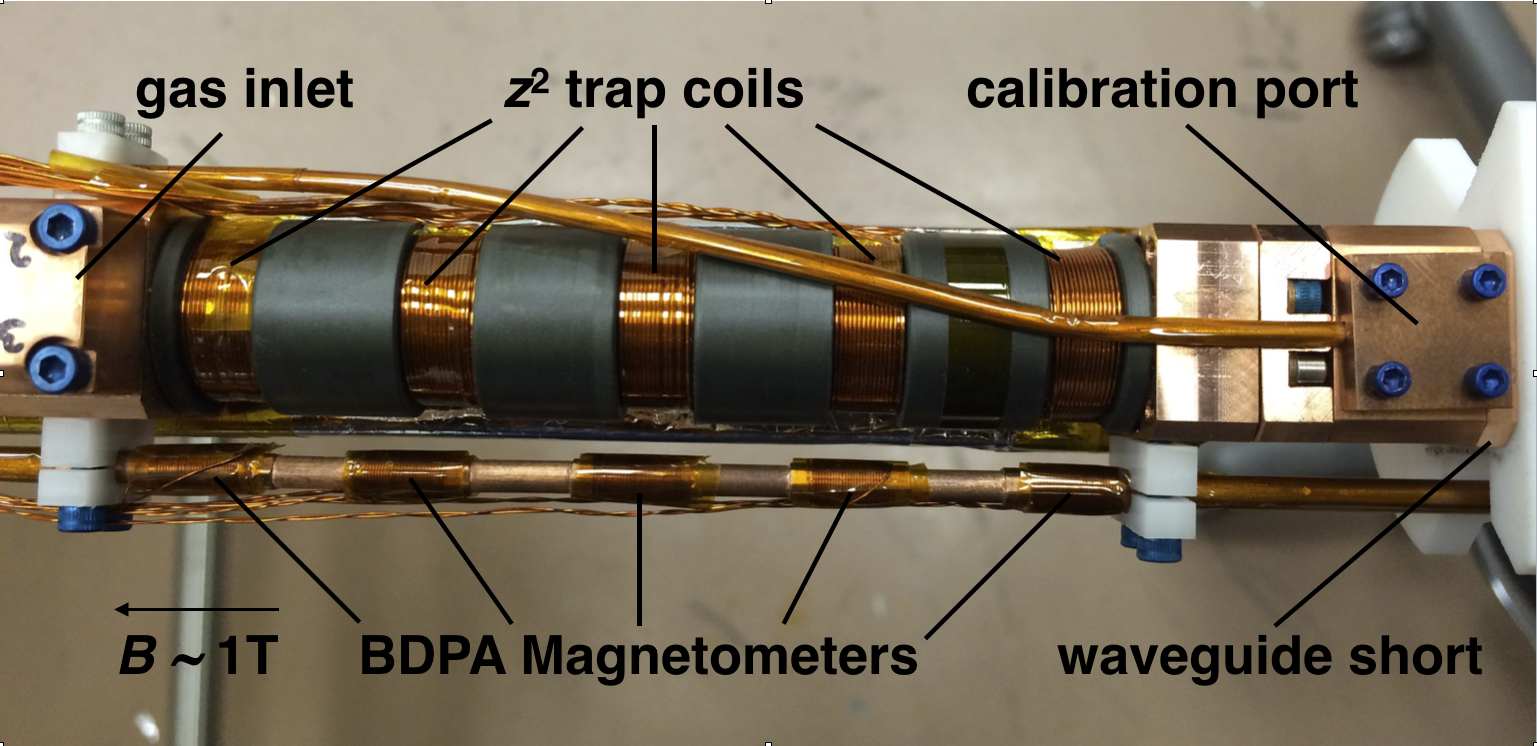}
  {CRES cell for Phase-II, showing the waveguide, trapping coils and
  magnetometers\label{fig:phase-II-cell}}

The cell shown in Figure~\ref{fig:phase-II-cell} differs from Phase I in having
a circular cross section that supports the propagation of circularly polarized
radiation. This configuration increases the effective volume and the
signal-to-noise ratio.  The magnetic trap consists of five copper coils, which can
act independently as harmonic traps, or in concert to form a broad flat region
with pinches at the end in the {\it bathtub} configuration. Each coil is closely paired with an Electron Spin Resonance (ESR) magnetometer for relative field strength measurements. The Phase I magnetometers, which used 2,2-diphenyl-1-picrylhydrazyl (DPPH), are replaced with magnetometers employing $\alpha$,$\gamma$-bisdiphenylene-$\beta$-phenylallyl (BDPA) to provide improved resolution in Phase II.

A circular-to-linear polarizer (``quarter-wave plate'') matches the cell to the WR-42 waveguide that
transports signals to the amplifiers. An additional feature is the insertion of
an isolator (a circulator with one terminated port) at the top of the waveguide.
The amplifiers have a VSWR (voltage standing-wave ratio) greater than one, and the isolator
provides a pure resistive termination at the amplifier temperature, about 35 K,
which flattens the noise spectrum. The cell is designed
with a copper body and CaF$_2$ windows to provide containment for tritium while
allowing low-loss transmission of microwave power. Non-evaporable getters store
the tritium and maintain an equilibrium pressure dependent on the selected
heating power.

As with the proof-of-concept apparatus, initial measurements are underway with
\isotope{Kr}{83 \rm m}, and the transition to tritium measurements will follow
once the performance of the system has been fully characterized.

\subsection{Phase III: Large-volume demonstrator\label{sec:phaseIII}}

Phase III evolves from previous phases by moving to a larger volume, which requires a
larger magnet and new methods for harvesting the microwave signal using a phased array of antennas.  With the larger instrumented volume we can achieve a neutrino mass sensitivity $m_\nu < 2$\un{eV}, competitive with current limits from Mainz~\cite{Kraus:2005fv} and Troitsk~\cite{Aseev:2011fk}. The tritium source will still be molecular. From Figure~\ref{fig:P8sensitivity} we see that such a limit could be attained in about 1 year of running with a source density $3\times10^{12}$\un{T_2/cm^3} in an effective volume of $10^{-5}$\un{m^3}$ = 10$\un{cm^3}.  If we assume 5\% electron trapping efficiency (similar to the trapping efficiency in Phases I and II, based on electron-tracking simulations), this limit requires 200\,cm$^3$ of physical volume, a contained activity of $7\times 10^{5}$ Bq and a mean track duration of 12$\un{\mu s}$. It is therefore not practical to conduct this experiment in an enclosed waveguide detector as in Phases I and II.  For Phase III, Project~8 must enlarge its tritium volume so that trapped electrons will emit cyclotron radiation into free space. A used MRI magnet has been purchased to accommodate the larger experiment, with an open bore of $90\un{cm}$.

A phased array of antenna elements is an effective means to collect free-space radiation.  We are investigating one or more ring-shaped arrays of antenna elements with each element amplified and digitized independently.  The area of the array focus in the plane of the array can be specified in post-processing by digital beam forming, in which the relative phases are adjusted in software before the signals from individual elements are combined~\cite{Turcotte:1999fk}.  The total available (coherent) signal power increases linearly with the number of instrumented channels $N$, while the incoherent noise of each channel contributes only $\sqrt{N}$ to the total noise~\cite{Ehyaie:2011fj}.  The antenna elements in current calculations are open-ended waveguides, but there are several other viable alternatives under consideration: resonant patches, Vivaldi (tapered slot), quasi-Yagi, and monopole elements. The antenna element used in the array must receive the linear polarization emitted by the electron as viewed in the plane of cyclotron rotation.  Numerical simulations have shown that the focal region of a ring array is a small ($\sim\!1\un{cm}$) spot in the plane of the ring, extending up to several cm in the perpendicular direction, with the total length depending on the radius of the ring.  Figure~\ref{fig:PhaseIII_focus} shows a 6\un{dB} electric field distribution in the cardinal planes of an array operating at 26\un{GHz}.

\twofigone[H]
  {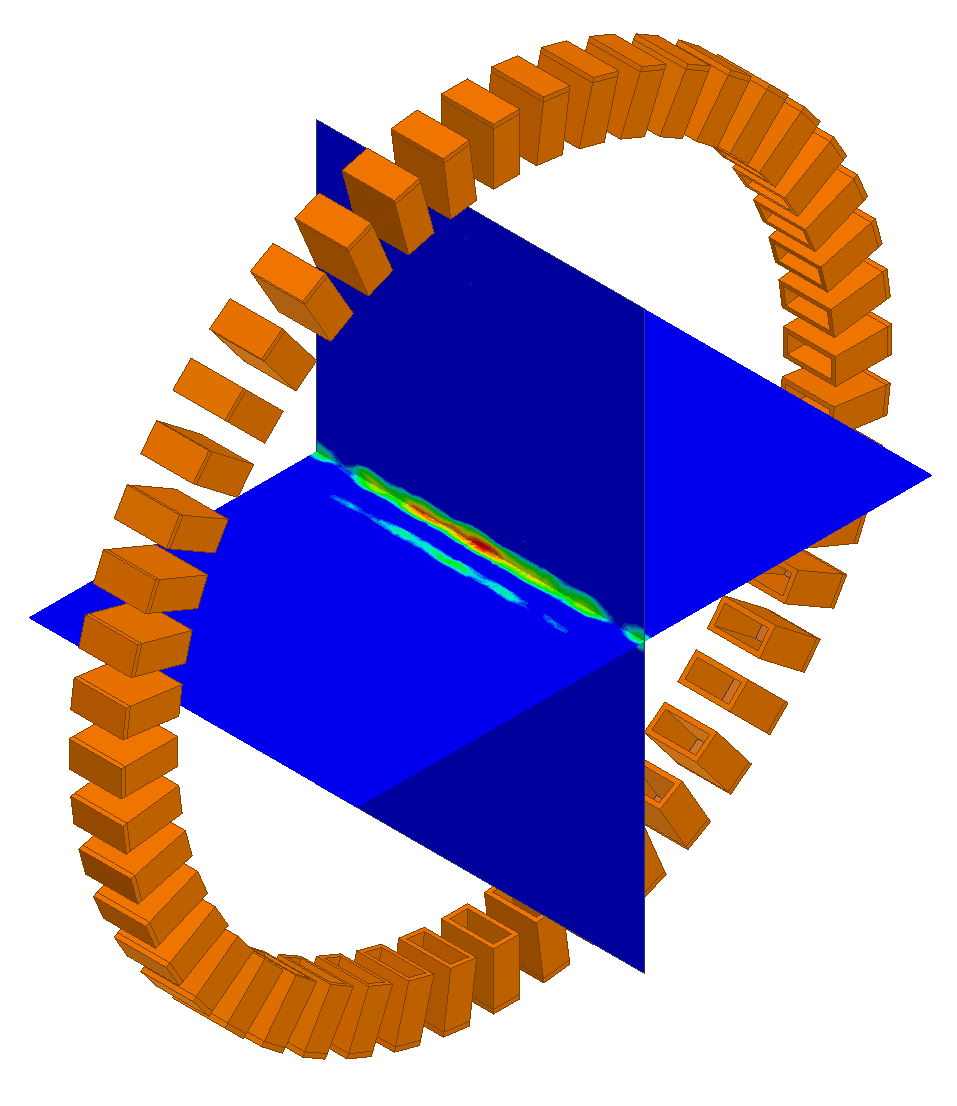}
  {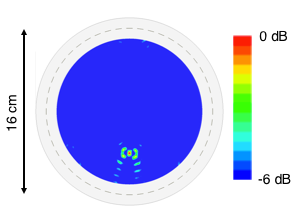}
  {The $6$-dB focal region (colored vertical and horizontal cut planes left and orthogonal projection right) of an 8-cm-radius ring array composed of 48 open-ended waveguide antenna elements in free space. Phases are tuned such that the focus is on the cylindrical axis (left) or 4\un{cm} from the center(right).\label{fig:PhaseIII_focus}}

A SNR of 9\un{dB} can be attained with an 8-cm-radius ring with 48 channels, the maximum that will fit around the circumference.  The area of the circular region shown in blue in~\ref{fig:PhaseIII_focus} is approximately 150\,cm$^2$, and the axial dimension is a few cm long, resulting in an instrumented volume larger than the 200\un{cm^3} physical volume required to attain a competitive limit in a year.

\subsection{Phase IV: Atomic tritium experiment\label{sec:phaseIV}}
The goal of Phase IV is sensitivity to the full range of neutrino masses allowed by the assumption of an inverted mass hierarchy.  To circumvent the fundamental limit set by final-state broadening with a molecular  T$_2$ source, Phase IV makes use of atomic tritium.  An idealized estimate of Project~8 sensitivity is $m_\nu \lesssim 40$\un{meV} (90\% C.L. - see Section~\ref{sec:sensitivity}).  Inspection of Figure~\ref{fig:P8sensitivity} shows that the effective volume of the atomic tritium source must be at least 10\,m$^3$, which requires 200\,m$^3$ of actual physical volume under the same assumption of 5\% electron trapping efficiency from Phase III.  Currently we do not have a complete design concept for Phase IV, and many aspects of this phase of the experiment will be challenging.  We can begin by considering the requirements of the atomic tritium source, from which we can derive some basic parameters that start to define the effort required to execute Phase IV.

We have made a semi-quantitative conceptual design of an atomic tritium source appropriate for Project~8~\cite{Clark:2014lq}.  The most obvious constraint it addresses is the maintenance of tritium in an atomic state (T), rather than its preferred molecular state (T$_2$).  The beta endpoint energy for T$_2$ is higher by 8\,eV than that of T, because of the higher mass of the daughter~\cite{Bodine:2015aa}.  The number of events in an interval $\epsilon$ below the endpoint is proportional to $\epsilon^3$.  Therefore even small contamination by an isotopologue with a higher endpoint introduces a significant background, affecting both statistical sensitivity according to Equation~(\ref{eqn:statistical_sensitivity}), and systematic sensitivity following from its energy dependence.  It is found that a relative purity of T$_2$/T $\lesssim 10^{-6}$ is required.  At vacuum pressures required for a precise CRES measurement, virtually all recombination to molecules occurs on the walls of the vessel.  Our solution is therefore to use magnetic confinement to keep the T, which has a magnetic moment, from contacting the vessel walls. T$_2$, with no magnetic moment, will rapidly leave the magnetic trap.  A magnetic field that confines the tritium will also trap the electrons under observation.  An appropriate magnetic field geometry is a Ioffe trap like the one used to trap anti-hydrogen in the ALPHA experiment~\cite{Amole:2014kx}.  A Ioffe trap has large gradients near the cylindrical walls that carry counter propagating axial currents, and negligible field far from the walls.  We find that a Ioffe trap 5\,T deep will confine atomic tritium at 130--170\,mK.  If the Ioffe trap has a 20-fold symmetry of current pairs then the ratio of fiducial volume to total vessel volume is 48\%.  The ratio increases to 75\% for 50-fold symmetry.  There is then the problem of source self-heating due to the high beta activity and scattering of electrons in the source.  A gas of \isotope{He}{4} (no magnetic moment) could be used to maintain thermal contact between the atomic tritium source and the cryogenic walls of a physical vessel.  The mean free path for He-T$^+$ scattering at vacuum pressures assumed for Phase IV is 50\,cm.  This sets a lower limit on the smallest physical dimension of the T source; tritium would evaporate from a smaller source due to insufficient cooling.  Targeted R\&D programs have begun within the Project~8 collaboration to address the experimental challenges.

\section{Summary and outlook}
\onefig[H]{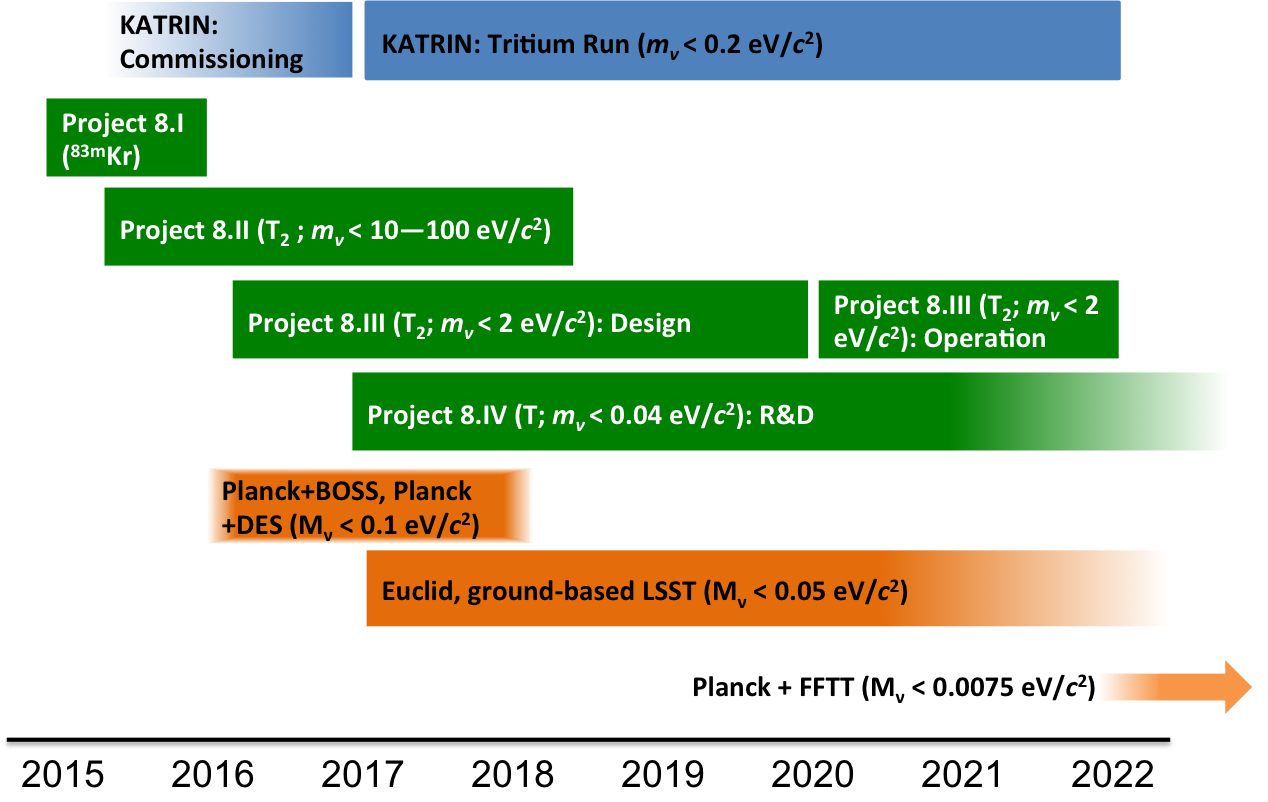}
 {\label{fig:P8Timeline} Timeline for Project~8 phases, with reference to expectations from KATRIN and cosmological observations.  Cosmological expectations are from Lesgourgues and Pastor~\cite{Lesgourgues:2014fk}}
 
The main challenges in determining the absolute neutrino mass from tritium endpoint spectroscopy arise from the very low event fraction within the narrow energy range where the effect of a non-zero neutrino mass can be observed, in combination with the required energy resolution, which is systematically limited by the excitation states of the tritium molecule. The newly developed Cyclotron Radiation Emission Spectroscopy (CRES) method has successfully resolved \isotope{Kr}{83m} lines at the \un{eV}-level in a small-volume waveguide setup. This motivates the development of a dedicated experiment using CRES to establish the absolute neutrino mass with a sensitivity down to $40\un{meV}$ when using atomic tritium as a source. The Project~8 collaboration pursues this goal in a staged approach, which individually addresses the main challenges: demonstration of the technique for tritium, a sufficiently large observation volume and an atomic tritium source. While numerous technical challenges and orders of magnitude in scale remain on this path, initial ideas for solutions are under development and major parallel efforts are underway on Phases I--III. 
Figure \ref{fig:P8Timeline} shows the projected timeline, in which we anticipate results that are competitive with both the existing MAC-E filter approaches and cosmological constraints.

\section*{Acknowledgments}
This material is based upon work supported by the following sources: the University of Washington Royalty Research Foundation; the Massachusetts Institute of Technology (MIT) Wade Fellowship; the U.S. Department of Energy Office of Science, Office of Nuclear Physics under Award No.~DE-FG02-97ER41020 to the University of Washington, under Award No.~DE-SC0011091 to MIT, under Award No.~DE-SC0012654 to Yale University, and under the Early Career Research Program to Pacific Northwest National Laboratory (PNNL), a multiprogram national laboratory operated by Battelle for the U.S. Department of Energy under Contract No.~DE-AC05-76RL01830; the National Science Foundation under Award Nos.~1205100 and 1505678 to MIT; and the Laboratory Directed Research and Development Program at PNNL.  A portion of the research was performed using PNNL Institutional Computing at Pacific Northwest National Laboratory. The isotope(s) used in this research were supplied by the United States Department of Energy Office of Science by the Isotope Program in the Office of Nuclear Physics. We further acknowledge support from Yale University, the PRISMA Cluster of Excellence at the University of Mainz and the KIT Center Elementary Particle and Astroparticle Physics (KCETA) at the Karlsruhe Institute of Technology.


\section*{References}

\bibliographystyle{iopart-num}
\bibliography{papers}

\providecommand{\newblock}{}
\begin{thebibliography}{10}
\expandafter\ifx\csname url\endcsname\relax
  \def\url#1{{\tt #1}}\fi
\expandafter\ifx\csname urlprefix\endcsname\relax\def\urlprefix{URL }\fi
\providecommand{\eprint}[2][]{\url{#2}}

\bibitem{Tribble:2007fk}
Tribble R {\em et~al.\/} 2007 {\em The Frontiers of Nuclear Science: A Long
  Range Plan\/} (Department of Energy and the National Science Foundation) \\
  \url{https://arxiv.org/pdf/0809.3137v1.pdf}

\bibitem{Geesaman:2015aa}
Geesaman D {\em et~al.\/} 2015 {\em Reaching for the Horizon: The 2015 Long
  Range Plan for Nuclear Science\/} (Department of Energy and the National
  Science Foundation) \\
  \url{https://science.energy.gov/~/media/np/nsac/pdf/2015LRP/2015_LRPNS_091815.pdf}

\bibitem{Planck:2016}
{Planck Collaboration} {\em et~al.\/} 2016 {\em A\&A\/} {\bf 594} A13 \\
  \url{http://dx.doi.org/10.1051/0004-6361/201525830}

\bibitem{Lesgourgues:2014fk}
Lesgourgues J and Pastor S 2014 {\em New Journal of Physics\/} {\bf 16} 065002
  \\ \url{http://stacks.iop.org/1367-2630/16/i=6/a=065002}

\bibitem{Alpert:2015}
{Alpert, B} {\em et~al.\/} 2015 {\em Eur. Phys. J. C\/} {\bf 75} 112 \\
  \url{http://dx.doi.org/10.1140/epjc/s10052-015-3329-5}

\bibitem{Gastaldo:2014}
Gastaldo L {\em et~al.\/} 2014 {\em Journal of Low Temperature Physics\/} {\bf
  176} 876--884 ISSN 1573-7357 \\
  \url{http://dx.doi.org/10.1007/s10909-014-1187-4}

\bibitem{Otten:2008zz}
Otten E~W and Weinheimer C 2008 {\em Rept. Prog. Phys.\/} {\bf 71} 086201 \\
  \url{http://dx.doi.org/10.1088/0034-4885/71/8/086201}

\bibitem{Angrik:xw}
Angrik J {\em et~al.\/} (KATRIN) 2005 {\em FZKA Scientific Report\/} {\bf 7090}
  \\ \url{https://publikationen.bibliothek.kit.edu/270060419}

\bibitem{Hannestad:2005ey}
Hannestad S 2006 {\em Prog. Part. Nucl. Phys.\/} {\bf 57} 309--323 [,309(2005)]
  \\ \url{http://dx.doi.org/10.1016/j.ppnp.2005.11.028}

\bibitem{Doe:2013fk}
{Doe} P~J {\em et~al.\/} 2013 {\em ArXiv e-prints\/} \\
  \url{https://arxiv.org/abs/1309.7093}

\bibitem{Asner:2015fk}
Asner D~M {\em et~al.\/} (Project 8 Collaboration) 2015 {\em Phys. Rev.
  Lett.\/} {\bf 114}(16) 162501 \\
  \url{http://dx.doi.org/10.1103/PhysRevLett.114.162501}

\bibitem{Monreal:2009za}
Monreal B and Formaggio J 2009 {\em Phys. Rev.\/} {\bf D80} 051301 \\
  \url{https://doi.org/10.1103/PhysRevD.80.051301}

\bibitem{Bodine:2015aa}
Bodine L~I, Parno D~S and Robertson R~G~H 2015 {\em Phys. Rev. C\/} {\bf 91}(3)
  035505 \\ \url{http://dx.doi.org/10.1103/PhysRevC.91.035505}

\bibitem{Kraus:2005fv}
Kraus C {\em et~al.\/} 2005 {\em Eur. Phys. J.\/} {\bf C40} 447--468 \\
  \url{http://dx.doi.org/10.1140/epjc/s2005-02139-7}

\bibitem{Aseev:2011fk}
Aseev V~N {\em et~al.\/} 2011 {\em Phys. Rev. D\/} {\bf 84}(11) 112003 \\
  \url{http://dx.doi.org/10.1103/PhysRevD.84.112003}

\bibitem{Lobashev:1985mu}
Lobashev V~M and Spivak P~E 1985 {\em Nucl. Instrum. Meth.\/} {\bf A240}
  305--310 \\ \url{http://dx.doi.org/10.1016/0168-9002(85)90640-0}

\bibitem{Picard:1992ys}
Picard A {\em et~al.\/} 1992 {\em Zeitschrift f{\"u}r Physik A Hadrons and
  Nuclei\/} {\bf 342}(1) 71--78 ISSN 0939-7922 10.1007/BF01294491 \\
  \url{http://dx.doi.org/10.1007/BF01294491}

\bibitem{Venos:2005vn}
V{\'e}nos D, Spalek A, Lebeda O and Fiser M 2005 {\em Applied Radiation and
  Isotopes\/} {\bf 63} 323--327 \\
  \url{http://dx.doi.org/10.1016/j.apradiso.2005.04.011}

\bibitem{Leber:2005aa}
Leber M 2005 private communications {\it Estimating the background rate for
  cosmic muons ionizing KATRIN source tritium molecules}

\bibitem{Carpenter:2015aa}
Carpenter B {\em et~al.\/} 2017 {\em Journal of Statistical Software\/} {\bf
  76} 1--32 ISSN 1548-7660 \\
  \url{https://www.jstatsoft.org/index.php/jss/article/view/v076i01}

\bibitem{Stan:2015aa}
{Stan Development Team,} 2015 Pystan: the python interface to stan, version
  2.9.0 \\ \url{http://mc-stan.org/interfaces/pystan.html}

\bibitem{Turcotte:1999fk}
Turcotte R, Ma S and Aguirre S 1999 Method and intelligent digital beam forming
  system with improved signal quality communications uS Patent 5,856,804 \\
  \url{https://www.google.com/patents/US5856804}

\bibitem{Ehyaie:2011fj}
Ehyaie D 2011 {\em Novel Approaches to the Design of Phased Array Antennas\/}
  Ph.D. thesis The University of Michigan \\
  \url{http://hdl.handle.net/2027.42/89713}

\bibitem{Clark:2014lq}
Clark B~M 2014 {\em Magnetic Trapping of Atomic Tritium for Neutrino Mass
  Measurement\/} Master's thesis California Institute of Technology \\
  \url{http://resolver.caltech.edu/CaltechTHESIS:07252014-082021105}

\bibitem{Amole:2014kx}
Amole C {\em et~al.\/} 2014 {\em Nuclear Instruments and Methods in Physics
  Research Section A: Accelerators, Spectrometers, Detectors and Associated
  Equipment\/} {\bf 735} 319 -- 340 ISSN 0168-9002 \\
  \url{http://www.sciencedirect.com/science/article/pii/S0168900213012771}

\end{thebibliography}
\end{document}